# Spectral dissociation of lateralized pairs of brain rhythms

Emmanuelle Tognoli[1]* & J. A. Scott Kelso[1,2]

[1] The Human Brain and Behavior Laboratory, Center for Complex Systems and Brain Sciences, Florida Atlantic University, Boca Raton, FL, USA

[2] Intelligent System Research Centre, University of Ulster, Derry, N. Ireland

Correspondence:
Emmanuelle Tognoli
Center for Complex Systems and Brain Sciences
Florida Atlantic University
777 Glades Road
Boca Raton, FL-33431, USA
Email: tognoli@ccs.fau.edu

Acknowledgements: This work was supported by grants from the National Institute of Mental Health (MH080838), the National Science Foundation (BCS0826897), the US Office of Naval Research (N000140910527), and the Davimos Family Endowment for Excellence in Science.

**Abstract**

Using high resolution spectral methods to uncover neuromarkers of social, cognitive and behavioral function, we have found that hemi-lateralized pairs of oscillations such as left and right occipital alpha, or left and right rolandic mu dissociate spectrally. That is, they show a shifted frequency distribution, with one member of the pair peaking at a slightly lower frequency than the other. To illustrate the phenomenon, we analyze EEG spatio-spectral patterns providing examples of dissociations in the 10Hz frequency band. Our observations suggest that homologous pairs of neuromarkers have distinct intrinsic frequencies and only transiently coordinate their oscillations into synchronous ensembles. On occasion, hemi-lateralized pairs are observed to blend into medial aggregates, leading to strongly coherent dynamics. We hypothesize that spectral dissociation plays a role in the balance of integration and segregation in the brain: separation of oscillations from homologous cortical areas allows them to function independently under certain circumstances, while preserving a potential for stronger interactions supported by structural and functional symmetries. As a method, spectral dissociation may be harnessed to better track the individual power of each member of a hemi-lateralized pair, and their respective time-courses. Resulting insights may shed light on the functional interaction between homologous cortices in studies of attention (alpha), e.g. during perceptual and social interaction tasks, and in studies of somatomotor processing (mu), e.g. in bimanual coordination and brain computer interfaces.

**Introduction**

Vertebrate organisms possess a striking, though incomplete symmetry in brain structural and functional organization (Braitenberg, 1977; Hugdahl, 2005; Halpern et al., 2005; Joliot et al., 2012). To a considerable extent, this may be determined by the (partial) decussation of sensory afferents and motor efferents. Such a scheme bisects the world in two roughly equivalent streams that are connected to each side of the brain, left cortex connecting predominantly with the right side of the world and vice versa. As a consequence, a number of brain areas have substantial left-right symmetry in both morphological configuration and connectivity. In turn, anatomical similarities may support similar functional behaviors, as seen in the neural oscillations from left and right homologous cortices in health and disease (Oostenveld et al., 2003; Goldensohn, 2007). Oscillations arise when brain areas engage in locally coherent activity (Bressler & Tognoli, 2006; Buzsaki, 2009). They possess characteristic frequencies and spatial organization: in particular, hemi-lateralized pairs of oscillations such as rolandic mu or parieto-occipital alpha rhythms have been recognized (Pfurtscheller & Berghold, 1989; Rodin & Rodin, 1995; Baboloni et al., 1999; Makeig et al., 2004). At a larger spatial scale, local oscillations are amenable to coordination with one another through the mechanisms of phase-locking and metastability (Kelso, 1995; Bressler & Kelso, 2001; Kelso & Tognoli, 2007). Such coordinated behavior provides further insight into the functional interactions of left and right homologous cortices.

In the following, we track left and right –lateralized oscillations and examine their coupled or independent expressions. To do so, we study multi-electrode spectra of sample human waking EEG, and compare the spectral distribution from homologous pairs of oscillations in the left and right sides of the brain. Perfectly overlapping spectral distributions (coming from oscillations that spend all of their time at the same frequency) hint strongly of sustainably phase-locked dynamics between left and right cortices. On the other hand, partially overlapping spectral distributions suggest that oscillations are only transiently coupled, whereas at other times, they function more independently. In the present report we observe that over the regular course of brain function, homologous oscillations overlap incompletely: they often exhibit separation of their spectral distributions and characteristic frequencies, suggesting that they operate with a degree of independence with respect to each other. Occasionally, homologous oscillations are seen to integrate their functional behavior, for example when a midline aggregate appears in lieu of left and right-lateralized oscillations, or when the spectral distribution from both sides of the brain completely overlaps. We discuss the implications of this dual tendency for integration and segregation, from the standpoint of the conceptual and empirical framework of Coordination Dynamics. Integration is a means by which a complex system uses its many parts cooperatively, thereby achieving a more efficient (collective) behavior than if its parts were acting independently (Kelso, 1995; Perez-Velazquez, 2009). Much evidence suggests that the brain achieves integration through phase-locking or metastable dwelling (Kelso, 1995; Varela et al., 2001; Uhlhaas et al., 2009; Tognoli & Kelso, 2009; Wang, 2010). Too comprehensive enslavement into collective behavior is deleterious to performance and complexity (Tononi et al., 1994; Kelso, 1995; Lenhertz, 2008; Tognoli & Kelso, 2013a). This limiting condition calls for a complementary concept (Kelso & Engstrom, 2006), segregation for this matter. Segregation expresses the autonomous behavior of the parts, and opposes to the formation of a collective behavior. The ability of the parts to segregate rests in large part on their broken symmetry, that is, inherent differences in their intrinsic properties or in their links with each other. Less-than-perfect symmetry in homologous brain rhythms may provide the originating structures with extended coordinative capabilities: at times, they are able to work with their homologous pairs, and at other times, they are able to free themselves to engage in other neural coalitions.

**Material and Methods**

Archival human waking EEG from our laboratory was reexamined (92 subjects) to detect bilateral variants of alpha and mu rhythms. Data encompassed mid- and high-density EEG (60-120 channels), sampled in a variety of tasks such as social coordination, bimanual coordination, movement-arrest, imitation, visual and auditory perception and rest (Tognoli et al., 2007; Tognoli, 2008; Tognoli et al., 2011a,b; Banerjee et al, 2012; for details). All EEG samples were acquired using a Neuroscan Synamps II (Compumedics, TX, El Paso). Data were collected from silver-silver chloride electrodes, referenced or re-referenced to a pair of electrodes positioned bilaterally at the loci of the mastoid processes. Prior to storage, all analog traces were filtered at 0.1-100Hz or 0.1-200Hz, and digitized at 1000Hz with 24bit vertical resolution in the range ±950µV. Resulting digital EEG was epoched and multichannel EEG spectra obtained. The spectrum of each channel was estimated using the Fast Fourier Transform (FFT) in long temporal windows (>8 seconds) to achieve a target spectral resolution greater than 0.12Hz/bin, which in our experience, provides satisfactory separation of EEG rhythms. Prior to the FFT, baseline corrected EEG epochs were convolved with a Tukey window (in steady-state tasks, -e.g. continuous finger movement or perception of rhythmic stimuli-, lasting more than 8.2 sec) or with a half-overlapping Hann window (for shorter steady-state tasks or for the study of transiently evoked responses). The latter segments were then zero-padded to achieve the target spectral resolution. Finally, multielectrode spectra were visualized using the colorimetric mapping introduced by Tognoli et al., (2007) to reveal spatio-spectral organization (Figure 1B). In this technique, the color assigned to the spectra from different channels is not arbitrary but inherited from electrode locations mapped onto perceptual distance, so that color similarity can be interpreted as a spatial property. For observers with normal color vision, this scheme helps to parse the spectra's spatial organization. Combined with high-spectral resolution, colorimetric mapping allows identification of multiple peaks, even when they overlap. Peaks are interpreted as local oscillations and their compounds. Peak frequencies and topographical distributions were used to identify rolandic mu (often seen in the upper range of the 10Hz band at electrodes C3 and C4,, i.e., above the central sulcus about 5-6cm left and right of the vertex in the 10 percent electrode coordinate system of Chatrian et al., 1985) and alpha rhythms (in the center of the 10Hz band near electrodes PO3/7 and PO4/8). We also examined oscillations that appeared at the spatial intersection of left and right alpha rhythms, the so-called medial alpha. Complementary analyses were conducted in the time-domain to study how oscillations transiently enter into collective ensembles. After suitable bandpass filtering of the EEG onto frequencies of interest, oscillations were visualized over time, and their phase aggregates identified (oscillations sampled from channels with leading amplitude for a given phase). When more than one oscillation is present in the study of phase aggregates (Tognoli & Kelso, 2009), the coordination dynamics of oscillations is examined in a pairwise manner. This is accomplished through the relative phase, computed as a difference between the oscillations' instantaneous phases (e.g. estimated from the Hilbert transform), wrapped over a modulo of 2 pi (one oscillation's cycle). Perpetuation of the same value of the relative phase (horizontal trajectory) suggests phase-locking indicative of integrative tendencies, and leading to closeness on the power spectrum; whereas change in the relative phase (ascending or descending trajectories) suggests escape into segregative behavior –which also leads to apartness in the power spectrum.

**Results and Discussion**

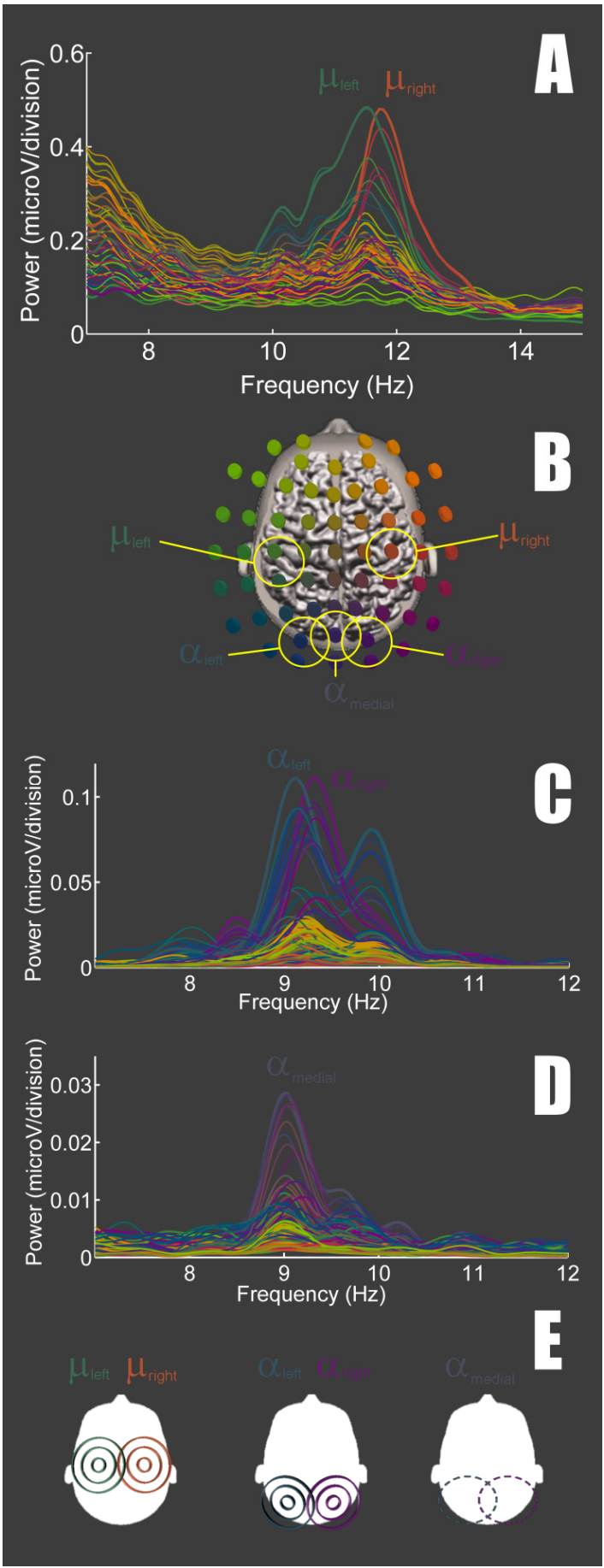

In a vast number of subjects, we observed a small but pervasive spectral separation of hemi-lateralized brain rhythms, as exemplified in Figure 1 which shows left and right rolandic mu (A) and left and right parieto-occipital alpha (C). In the sample shown in Figure 1A and representative of tasks that engage and disengage motor behavior, two spectrally distinct mu rhythms are revealed, as can be seen by looking at the green and orange peaks appearing about 11-12Hz. The peaks have a small but clear spectral separation (about 0.23Hz apart in this subject, a separation made visible by the high spectral resolution that we employed, here 0.03Hz). The alpha rhythms are generally manifest in tasks that engage and disengage attentive vision. The pair shown in Figure 1C has a more complicated multi-peak representation than the one shown in sample (A), but again exhibits distinct spectral distributions, as seen from the intermingling blue and purple traces. Sample spectra are also observed when only one of the two homologous rhythms is present (not shown: for instance only left but not right mu, or only right but not left alpha), suggesting that homologous rhythms need not be co-present at all. Together, these observations suggest that homologous rhythms may possess a degree of independence from one another. This "room for independence" occurs against the odds of their rather similar anatomical connectivity which favors strong synchronization and identical distribution of spectral energy.

*Figure 1: Sample spectra showing dissociation of hemi-lateralized pairs of brain rhythms (A,C) as well as an example of their integrated counterpart (D). All spectra are encoded according to the colorimetric mapping shown in B. (A) reveals a lateralized pair of mu rhythms above rolandic regions (green and orange peaks and their respective spatial decay) in a representative subject. Note separation of left and right mu spectra over the frequency axis. Similarly, C shows a pair of parieto-occipital alpha rhythms (blue and purple peaks), also exhibiting a dissociation, albeit of a slightly more complex form. In other conditions, spectra from the same subject as in (C) show a distinct spatial organization (D, mauve peak), suggestive of a medial aggregate: we hypothesize that previously independent left and right alpha merge spatially and spectrally to form a new peak at their spatial intersection. (E) recapitulates the spatial inferences attributed to the 5 observed rhythms. Left and right mu are interpreted as local oscillations arising from two lateralized somatomotor areas. Left and right alpha are interpreted as arising from a pair of parieto-occipital areas, whereas medial alpha is attributed to their intersection. Further details in text.*

This partially independent spectral distribution of left and right homologous rhythms is not, however, a constant characteristic. Figure 1D shows spectra from the same subject as in Figure 1C, though sampled on a different occasion and in another experimental condition. A midline rhythm over parieto-occipital regions is seen (mauve color located above the longitudinal fissure), exactly at the intersection of previously shown left and right alpha. Further, the peak shown in figure 1D has a frequency distribution that matches quite well the subject's left and right parieto-occipital alpha from figure 1C. Such combined spatial and spectral coincidences weaken the interpretation that the neuromarker from figure 1D is distinct from the previously mentioned pair. Instead it leads us to hypothesize that the peak seen in figure 1D is a compound of left and right parieto-occipital alpha (illustrated in figure 1E right), that is, the outcome of a pair of lateralized rhythms whose respective electrical fields combine into a midline aggregate. Source estimation could assist in testing this hypothesis, though it would have to deal efficiently with sources presenting a strong spectral overlap (Murzin et al., 2011). In contrast with the pairs of rhythms shown in figures 1A and C, the medial alpha compound (D) only appears if its contributing left and right alpha components have completely overlapping spectra, as may happen when cortices engage in phase-locked behavior. We therefore propose that the medial alpha rhythm is seen in instances when left and right alpha rhythms engage in strongly coherent behavior (spatial orientation of the alpha generators may also contribute to the phenomenon). In such a situation, it may not be possible to track the power of each neuromarker independently, though it remains crucial to establish precise temporal boundaries to quantify episodes of phase-locking and metastable dwelling.

Unfolding the temporal dynamics of neural oscillations enables one to conduct a detailed analysis of the rhythms' coordination dynamics (Tognoli & Kelso, 2009). The goal is to determine when and how oscillations are governed by integrative tendencies and when they are not. This is illustrated in figure 2, with the example of left and right mu rhythms (upper plot, bold oscillations in green and orange, respectively, accompanied with whole scalp traces for spatial context). In this sample lasting about 26 cycles over 2.5 seconds, moments are shown which reveal only left mu (before annotation A); only right mu (after annotation B); or both left and right mu (between annotations A and B, and seen through various modes of coordination). Over the period when left and right mu coexist, their coordination dynamics (lower plot of figure 2) reveals alternation of phase wrapping (non-horizontal segments of the relative phase indicative of segregative tendencies) and phase dwelling (horizontal segments of the relative phase indicative of integrative tendencies, for instance the long antiphase episode lasting about 7 cycles marked in red). The changing coordination dynamics of left and right mu is typical and likely to carry information through the patterning of relative phase (Bressler & Kelso, 2001; Varela, et al., 2001; Kelso & Tognoli, 2007; Palva & Palva, 2012).

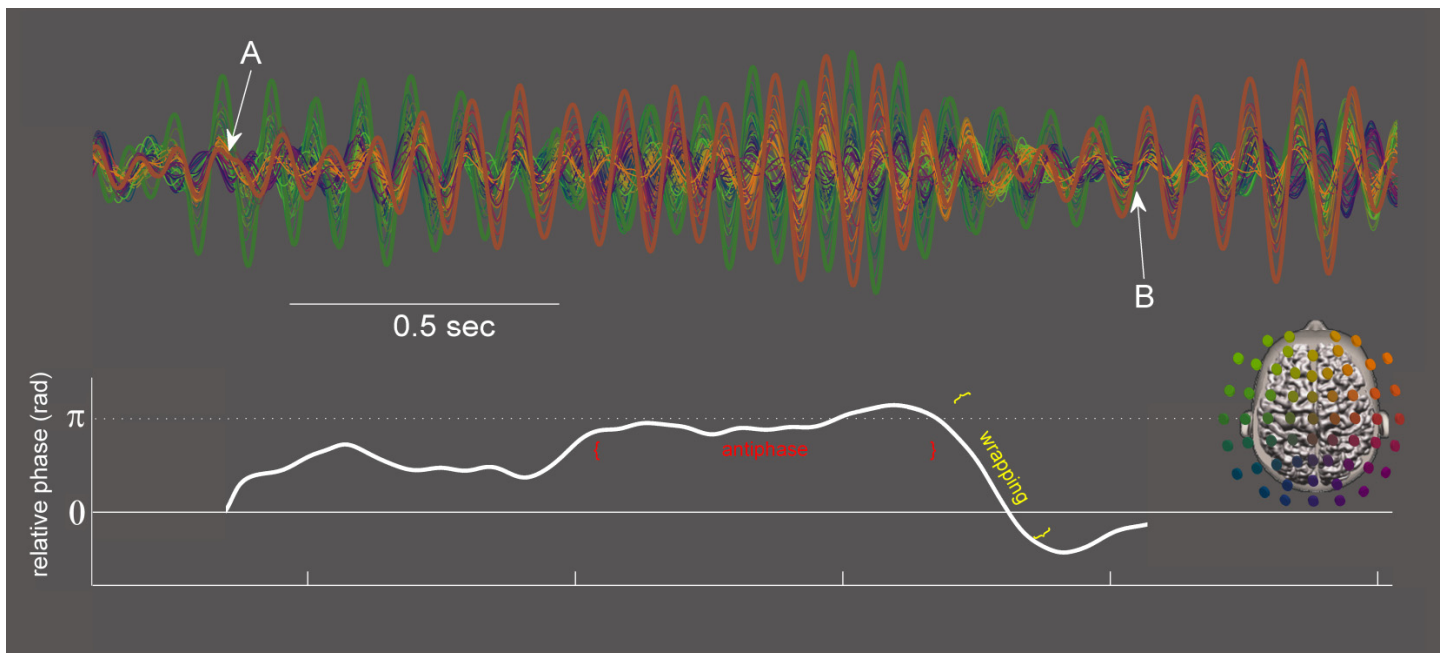

*Figure 2: Sample oscillations from bandpass-filtered EEG showing the integrative and segregative tendencies of left and right mu rhythms over 2.5 sec (top) and their coordination dynamics through the relative phase (bottom). Left mu is emphasized by the bold green trace, and right mu with the bold orange one. All other channels are displayed for spatial context, encoded with colors as shown on the colorimetric legend (lower right). On the top figure, a changing coordination behavior is observed, starting with the presence of only left mu (to annotation A), followed by the recruitment of right mu that initially stays segregated from left mu before dwelling into collective behavior, mostly near antiphase, and then followed by the removal of left mu (after annotation B). The coordination dynamics is accompanied by changes in amplitude, part of which is spuriously driven by phase bias (Tognoli & Kelso, 2009), with underestimation of antiphase and overestimation at inphase. When both mu are present, the coordination dynamics of their phases (bottom) shows a rich patterning involving a succession of segregative and integrative tendencies. See details in text.*

From the viewpoint of coordination dynamics, excessive symmetry promotes strong coupling between brain regions (Kelso, 2009). As both a cause and a consequence, brain regions cannot easily escape from one another to organize alternate coalitions. In a system with strong symmetries, effecting transitions during the course of brain function places high demands in terms of time and energy (Kelso & Tognoli, 2007). Such a system has less than optimal complexity and performance: the same goes for a system with too weak symmetries on the other end of the continuum. The observations reported here suggest that homologous pairs of brain rhythms such as mu and alpha possess substantial similarities (appreciated from their spectral proximity) as well as some degree of independence (supported by observations that homologous spectra do not fully overlap in their frequency distribution and midline aggregates do not appear systematically). We hypothesize that this design provides homologous brain areas with the flexibility to engage in and disengage from reciprocal coordination (Tognoli & Kelso, 2013b). The present observations lay some foundations for empirical and theoretical scrutiny on the functional interactions between the hemispheres and may not be unique to the 10Hz frequency range. The overall anatomo-functional organization of the brain suggests that hemi-lateralized pairs are likely to exist in other frequencies, an hypothesis worth exploring.